\documentclass[copyright,creativecommons]{eptcs}
\providecommand{\event}{SecCo 2009}
\providecommand{\volume}{??}
\providecommand{\anno}{20??}
\providecommand{\firstpage}{2}
\providecommand{\eid}{??}
\usepackage{breakurl}

\usepackage{amsmath}
\usepackage[all]{xy}

\usepackage{graphicx, subfigure}

\newtheorem{definition}{Definition}[section]

\newtheorem{proposition}[definition]{Proposition}
\newtheorem{theorem}[definition]{Theorem}

 \newcommand{\myproof}[1]{{\em Proof}:  {#1 } \vspace{1cm}}

\title{Studying Maximum Information Leakage Using Karush--Kuhn--Tucker Conditions} 

\author{Han Chen
\institute{School of Electronic Engineering and Computer Science\\ Queen Mary University of London}
\email{hanchen@dcs.qmul.ac.uk}
\and
Pasquale Malacaria
\institute{School of Electronic Engineering and Computer Science\\ Queen Mary University of London}
\email{pm@dcs.qmul.ac.uk}
}
\def\titlerunning{Studying Maximum Information Leakage Using Karush--Kuhn--Tucker Conditions}
\def\authorrunning{Han Chen \& Pasquale Malacaria}
\begin{document}
\maketitle 

\begin{abstract}
When studying the information leakage in programs or protocols, a natural question arises: ``what is the worst case scenario?". This problem of identifying the maximal leakage can be seen as a channel capacity problem in the information theoretical sense. In this paper, by combining two powerful theories: Information Theory and Karush--Kuhn--Tucker conditions, we demonstrate a very general solution to the channel capacity problem. Examples are given to show how our solution can be applied to practical contexts of programs and anonymity protocols, and how this solution generalizes previous approaches to this problem.
\end{abstract}


\section{Introduction}

As emphasized in the existing literature, no electronic system can guarantee perfect confidentiality or anonymity \cite{2pas}. Hence,
measuring the leakage of  confidential information is a pressing but increasingly challenging issue. The ability to preemptively assess possible information leaks is crucial for designing and understanding a system that contains information which ought to be protected \cite{NAE}.  

Information Theory \cite{1sha} provides a general method for measuring information flow in information channels, and extends to quantify the loss of confidentiality and anonymity. A number of previous works have addressed and measured the channel capacity of information leakage channels, which describes the {\em worst-case} leakage. Recently a novel technique to measure the channel capacity of anonymity protocols and programs using Lagrange multipliers has been proposed in \cite{3pas,3han}: this setting is able to answer questions like:  ``what is the maximum leakage of a system where a random string is 1000 times less likely to be the secret than a dictionary word" i.e. an equality constraint like  ${\tt p_{rand}=1000 p_{word}}$.\footnote{By maximum leakage we mean the maximum number of bits leaked. Notice that this is different from the maximum percentage of the secret leaked.}

In order to analyze a much wider range of systems and scenarios, inequality constraints ought to be supported. An example of such constraint is: ``the password is over 1000 times more likely to be a word from a dictionary than a meaningless string", i.e. ${\tt p_{rand}<1000 p_{word}}$: these inequality constraints cannot be solved using lagrangians. Therefore, we introduce Karush--Kuhn--Tucker (KKT) conditions
to enable inequality constraints for deriving the channel capacity, and present a set of theorems and propositions which can be readily applied. 
This makes the approach more powerful and enables it to deal with a much wider spectrum of cases, as demonstrated later on in this paper. Further, we believe that this approach, orthogonal to the probabilistic methods which have dominated protocol security analysis \cite{FeJoSy,DiSeClPr,1ree}, will provide novel and more practical results to the research community.

The paper is organized as follows: the next subsection discusses existing literature and the background is introduced in Section \ref{sec:bg}. In Section \ref{cckkt}, we briefly describe the theorems and propositions for channel capacity using Karush-Kuhn-Tucker conditions with full proofs. We show that our method can be applied to programs and protocols in Section \ref{app}. Finally, we provide concluding remarks and discuss future works in Section \ref{clf}.

\subsection{Related Works}

This work extends from previous works by Chen and Malacaria \cite{3pas,3han}.
Information Leakage is measured using the same Information Theoretical definitions used by a number of authors\cite{0dav,2pas,1bor,kopf1,kopf2}, and
follows pioneering works by Denning\cite{1den}, Gray\cite{1gre}, Mclean\cite{1mcl} and Millen\cite{1mil}. 
A recent alternative Information Theoretical definition of leakage has been proposed by Smith \cite{1smi}  in terms of min entropy; in the context of protocols  those ideas have been investigated by \cite{BCP}. A discussion of the relation between min entropy and Shannon entropy relevant to the context of this work can be found in \cite{4pas}.
In a program analysis context channel capacity has been recently investigated in \cite{NeMcCaSong}. 

Channel capacity of anonymity protocols in a restricted context has been characterized by Chatzikokolakis, Palamidessi and Panangaden \cite{cpp1}. However, their method applies to protocols with ``symmetric" properties. These restrictions are overcome in \cite{3pas,3han} where Lagrange multipliers are used to compute the maximum leakage of deterministic programs and anonymity protocols with additional equality constraints.  
 Blahut \cite{1bla} mentioned KKT conditions while proposing his iterative algorithm for approximate channel capacity. However, the use of KKT conditions in the context of this work is original.

There is a large set of work on anonymity protocols using probabilistic techniques \cite{FeJoSy,DiSeClPr,1ree}. 
A probabilistic approach would assume a certain kind of distribution to work out an expectation of anonymity in a given model. In comparison, 
our method allows for the use of more flexible relationships to track the maximum leakage, which is a pressing problem that remains largely unsolved.  
Whilst it is known that information theoretical and probabilistic notions are related, the extent of this relationship requires further investigation.   

\section{Background}\label{sec:bg}
In this work we refer to a program or a protocol as an {\em information leakage channel}.
We define an information leakage channel as a triple
\[\langle { \cal H}, {\cal O}, \phi \rangle\]
where the input, ${\cal H}$, is a set of confidential information, and the output,
${\cal O}$, is a set of observations. To introduce probabilities we use two random variables: 
$h$ for $\cal H$ and $O$ for $\cal O$ respectively. We also denote members of $\cal H$ as $h_i \in \cal H$
, and members of  $\cal O$ as $o_j \in \cal O$. $\phi$ describes the conditional probability between the two random variables:
\[ \phi = P(O \vert h) \]
In deterministic channels, one input $h_i$ can only produce one output $o_j$ thus $\phi_{j, i} = 1$.

With this definition, both programs and anonymity protocols can be seen as 
information leakage channels. In general, information leakage channel 
has three elements: confidential information as inputs, public information as observations and 
the corresponding probabilities between them. 


The triple above: $\langle {\cal H},{\cal O}, \phi \rangle$ can be represented by a probability matrix: rows describe elements of $\cal H$, columns describe elements of $\cal O$ and the 
value at position $(h_i, o_j)$ is the conditional probability $\phi_{j,i}$. This is the chance of observing $o_j$ given $h_i$ as the input.

\subsection{Background on Karush--Kuhn--Tucker Conditions}\label{lan}

Karush--Kuhn--Tucker (KKT) conditions \cite{kkt} generalize the Lagrange method for finding the extrema of a function subject to a family of constraints: while Lagrange multipliers consider only equality constraints,  KKT conditions allow for general inequality constraints. We refer the reader to \cite{3pas,3han} for a background on Lagrange Multipliers in this context.

\subsection{A Simple Example}

We will illustrate the use of the method by a simple example below. 

Suppose we want to maximize the following function:
\[ f(x, y)=xy \]
subject to the inequality constraint
\[x+y \leq 8 \]
First we construct the Lagrange function which combines the original function waiting to be maximized and the constraint
\[L(x, y)=xy+\lambda(8-x-y)\]
where $ \lambda$ is a number which indicates the weight associated with the constraint, for example
ignoring the constraint is equivalent to setting $ \lambda=0$.

Formally, the term $\lambda$ which is the {\em Lagrange multipliers} and the Lagrange technique are used in order to find the maximum of the function by differentiating on $x,y$ {\em and} $\lambda$.

Using KKT we get the optimal solution for the original optimization problem by solving the following equations:

\[ \frac{\delta L(x, y)}{\delta x} = y-\lambda=0,  \ \  \
 \frac{\delta L(x, y)}{\delta y} = x-\lambda =0, \ \ \ 
\lambda(8-x-y)=0\]

We deduce the additional constraint
\[ x+y \leq 8, \ \  x=y=\lambda \]

We use the conclusion $x=y=\lambda$ to replace $x$ and $y$ in the original function and get the maximal value
\[  \textrm{Max}f(x,y)= \lambda^{2}\]

Notice there is another equation we didn't use so far $\lambda(8-x-y)=0$, from this and the constraint $x+y \leq 8$ we get two cases
\[ \lambda=0,\  \text{or} \ x+y=8\]
It's easy to see that $\lambda=0$ is a saddle point because this value can not give the local maximum. Then we use $\lambda$ replace the variables $x$ and $y$ in the second case and get
\[ 2\lambda=8 \Rightarrow \lambda=4\]

It is then easy to derive the values for the other variables i.e.
\[ x=4,\ \ y=4\]

Now the values $x=4,y=4$ do satisfy the constraint and also are the values that maximize the original function 
\[ \textrm{Max} f(x, y) = xy =16\] 

\subsection{Theoretical Basis of Karush--Kuhn--Tucker Conditions} \label{ILagT}

We consider the problem of finding the extrema of a function $f$ subject to a family of constraints $C_{1\leq i\leq m}$ where $C_i$ is an inequality of the form $ g_i(x) \geq b_i$.

The first step is to construct the Lagrange function $L(x,\lambda)$ where $\lambda$ is a Lagrange multiplier for the inequality constraint which is similar as the multiplier for equality constraint. 
The inequality constraints are expressed in the form  $g_i(x) - b_i \geq 0$ and then we introduce the $\lambda$ associated with the constraints.
 
In a general setting let $L(x,\lambda)$ be the Lagrangian of a function $f$ subject to a family of constraints $C_{1\leq i\leq m}\ (C_i\equiv g_i(x) \geq b_i)$, i.e.
\[  L(x,\lambda)=f(x)+\sum_{1\leq i\leq m} \lambda_i(g_i(x)-b_i) \]

The basic result justifying KKT method is the following:
\begin{theorem}\label{Ltheorem}
Assume the vector $x^{*}=(x^{*}_1,\dots,x^{*}_n)$ maximizes (or minimizes) the continuous function $f(x)$ subject to the constraints $(g_i(x) \geq b_i)_{1\leq i\leq m}$. Then either
\begin{enumerate}
\item the vectors $(\nabla g_i(x^{*}))_{1\leq i\leq m}$ are linearly dependent, or
\item there exists a vector $\lambda^{*}=(\lambda^{*}_1,\dots,\lambda^{*}_m)$ which is an optimal solution for the original optimization problem satisfying the following conditions
 \\$\nabla L(\lambda^{*},x^{*})~=~0$ \ \  i.e.
\[ (\frac{\delta L}{\delta x_i}(x^{*})=0)_{1\leq i\leq n} \]
and 
\[\lambda^{*}(g_i(x^{*})-b_i)=0, \ \ g_i(x^{*}) \geq b_i, \ \ \lambda_i \geq 0 \]
\end{enumerate}
\end{theorem}

\noindent where $\nabla$ is the gradient and these conditions are called KKT  conditions. 

The condition $\lambda_i \geq 0$ implies non-negative Lagrange multiplier and $\lambda^{*}(g_i(x^{*})-b_i)=0$ implies two cases:

\begin{eqnarray}
g_i(x^{*})=b_i\\
g_i(x^{*}) > b_i \rightarrow \lambda_i =0
\end{eqnarray}

\subsection{Results of Lagrange Multipliers: A Short Review}

We now give a short  review of the results  in \cite{3pas,3han}. These works  use Lagrange multiplier to solve the channel capacity in programs and anonymity protocols with equality constraints. 

\begin{theorem}\label{non-det-eq}
In probabilistic channels, the probabilities $h_i$ maximizing $I(h; O) $ subject to the family of constraint $(\mathcal{C}_{k})_{k\in K}$ are given
 by solving in $h_i$ the equations 
 \begin{eqnarray*}
 \sum_{o_s\in \hat{O_i}} \phi_{s,i} \ln( \frac{\phi_{s,i} } { o_s} )    -  1+ \sum_k \lambda_k f_{i,k}&=&0
\end{eqnarray*}
and the constraints  $(\mathcal{C}_{k})_{k\in K}$.
\end{theorem}
where $\phi_{s,i}$ is the probability of observing $o_s$ when the input is $h_i$; $\hat{O_i}$ denotes the set of observations possible for the secret $h_i$; $f_{i, k}$ is the factor of $h_i$ in the $k_{th}$ constraint.

Using the probabilities $h_i$ we can work out the channel capacity.

\begin{proposition}\label{maxleakage:chanCap}
The channel capacity is given by
\[ \sum_i h_i(1- \sum_k \lambda_kf_{i,k})d\]
\end{proposition}

If the system is deterministic, the  formula in Theorem \ref{non-det-eq}  can be simplified to 
\begin{eqnarray*}
\-\ln(o_i) -  1+ \sum_k \lambda_k f_{i,k}&=&0
\end{eqnarray*}

 Moreover, in the case of the single constraint $\sum_i h_i=1$ the channel capacity of deterministic information leakage channels can be further simplified.
\begin{proposition}\label{P2}
 The channel capacity of deterministic information leakage channels without any additional constraint is given by
\[  d(1-\lambda_0) \]
where $d=\frac{1}{\ln 2}$.
\end{proposition}
From Theorem \ref{non-det-eq} we can know that Proposition \ref{P2} implies the well known fact that the channel capacity of unconstrained deterministic programs is the log of the number of possible outputs.

\section{Channel Capacity using Karush--Kuhn--Tucker Conditions}\label{cckkt}

\subsection{Constraints}

Often the attacker's knowledge about the secret can be expressed in terms of inequalities: for example, ``a unix password is 100 times more likely to be a word from a dictionary than a meaningless string".
We hence need KKT conditions to compute channel capacity in this context.
Remember that there is always at least one constraint for the input distribution requiring that the sum of their probabilities is $1$; we denote this constraint as $C_0$. Additional constraints are used to specify the conditions of inputs needed to satisfy:  we use $C_k$ for these conditions.
\begin{eqnarray*}
C_0 \equiv \sum h_i=1\\
C_k \equiv g_k(h_i) \geq F_k \ \ (k > 0)
\end{eqnarray*}
\noindent where $F_k$ are constants and $g_k(h_i)$ are ``statistics" or expectations , i.e. linear inequality expressions in the form of 
\[g_k(h_i) \equiv \sum_i h_i f_{i,k}\]
KKT conditions only provides precise solutions for non strict inequalities;  for strict inequalities, we can only provide an approximate solution.

\subsection{Theory and Proof}

{\bf Convention:}

As previously explained, we denote $h_i$ as the $i$-th possible value that the variable $h$ can assume. Also, $o_j$ denotes the $j$-th possible value for the observation variable $O$.
Each possible event $h_i$  has a given probability $\mu(h_i)$. To ease the exposition we will use $h_i$ both for the event $h_i$ and for its probability $\mu(h_i)$, and similarly $o_j$ for $\mu(o_j)$. However when it is clear from the context we may use $h_i$ for the $i$-th value of the variable $h$, i.e. $h=v_i$. The context will disambiguate what meaning is intended. 

As usual we use the conditional probability of $\phi_{k, i}$ for the probability of observing $o_k$ given the input $h_i$. Using Information Theory we have:
\begin{eqnarray*}\label{fun}
I(h; O)&=& H(h)-H(h \vert O)\\
&=&  H(h)+ \sum_{k } o_k \sum_i (h_i \vert  o_k) \log(h_i \vert  o_k)\\
&=& H(h)+ \sum_{i, k} (h_i, o_k) \log(h_i \vert  o_k)\\
&=& -\sum_ih_i\log(h_i)+ \sum_{i,k} h_i\phi_{k,i} \log(\frac{h_i\phi_{k,i}}{o_k})\\
&=& - \sum_{i,k}h_i\phi_{k,i}\log(h_i)+\sum_{i,k} h_i\phi_{k,i} \log(\frac{h_i\phi_{k,i}}{o_k})\\
&=&  \sum_{i,k}h_i\phi_{k,i} \log(\frac{\phi_{k,i}}{o_k})\\
\end{eqnarray*}


Notice that 
\begin{eqnarray*}\label{fu}
 \sum_{i,k}h_i\phi_{k,i} \log(\frac{\phi_{k,i}}{o_k}) & =&  d \sum_{o_s\in \hat{O_i} } h_i\phi_{s,i} \ln{\frac{\phi_{s,i}}{o_s}} + \ d \sum_{o_s\in \hat{O_i},h_r \in P_i}  h_r\phi_{s,r}\ln(\frac{\phi_{s,r}}{o_s}) )   
 \end{eqnarray*}
where $d=\frac{1}{\ln2} $  and 
\[ P_i= \{ h_r  \vert  \phi_{s,r}\not=0,  o_s\in \hat{O_i},r\not= i \} \]
where in  the formula $\hat{O_i}$ denotes the set of observations possible for the secret $h_i$ (i.e. the set of non zero observations compatible with input $h_i$).

Assuming a set of constraints $(\mathcal{C}_k)_{k \in K} \equiv g_k(h_{i}) \geq F_k$, the Lagrange function hence becomes
\begin{eqnarray*}
L(h_i)&=& I(h;O)+d\sum_k \lambda_k (\sum_i h_i f_{i,k}-F_k)\\
&=&  d \sum_{o_s\in \hat{O_i} } h_i\phi_{s,i} \ln{\frac{\phi_{s,i}}{o_s}} + d \sum_{o_s\in \hat{O_i},h_r \in P_i}  h_r\phi_{s,r}\ln(\frac{\phi_{s,r}}{o_s}) ) + d\sum_k \lambda_k (\sum_i h_i f_{i,k}-F_k)
\end{eqnarray*}
 where $d=\frac{1}{\ln 2}$ is used to convert the logarithm in base 2 $\log$ into natural logarithm $\ln$.
 
 As mentioned earlier, we always assume the constraint
$C_0\equiv\sum h_i=1$.
 
Using KKT the maximum $L(h_i)$ is given by the following theorem:

\begin{theorem}\label{non-det-ieq}
In information leakage channels, the probabilities $h_i$ maximizing $I(h; O) $ subject to the family of constraint $(\mathcal{C}_k)_{k \in K} \equiv g_k(h_{i}) \geq F_k$ are given
 by solving in $h_i$ the following system of inequalities:
\[ \sum_{o_s\in \hat{O_i}} \phi_{s,i} \ln( \frac{\phi_{s,i} } { o_s} )    -  1+ \sum_k \lambda_k f_{i,k}=0 \land  \lambda_k \geq 0,\ \ g_k(h_{i}) \geq F_k \]
or
\[ \sum_{o_s\in \hat{O_i}} \phi_{s,i} \ln( \frac{\phi_{s,i} } { o_s} )    -  1+\lambda_0=0 \land g_k(h_{i}) \geq F_k \]
\end{theorem}

\myproof{
Recall that the KKT conditions are
\[ (\frac{\delta L}{\delta h_i}(h^{*})=0)_{1\leq i\leq n} \]
\[ \lambda_k(\sum_i h_i f_{i,k}-F_k)=0, \ \ \sum_i h_i f_{i,k} \geq F_k, \ \ \lambda_k \geq 0\]
Compared to the KKT conditions for equality constraints we found that there are three additional ones:
\[ \lambda_k(\sum_i h_i f_{i,k}-F_k)=0, \ \ \sum_i h_i f_{i,k} \geq F_k, \ \ \lambda_k \geq 0\]
other than
\[\frac{\delta L}{\delta \lambda_i}(\lambda^{*})=0_{1\leq i\leq n} \] which actually represents the constraints
\[\sum_i h_i f_{i,k}-F_k=0\]

Firstly we simplify the three additional constraints as

\begin{displaymath}
\begin{array}{c}
\lambda_k(\sum_i h_i f_{i,k}-F_k)=0 \\
\sum_i h_i f_{i,k} \geq F_k\\
\lambda \geq 0
\end{array}\Rightarrow 
\begin{array}{c}
\sum_i h_i f_{i,k}=F_k \land \lambda_k \geq 0  \  \text{or}\\
\sum_i h_i f_{i,k}\geq F_k \land \lambda_k =0 
\end{array} 
\end{displaymath}

Combine the result with the derivative condition 
\[ (\frac{\delta L}{\delta h_i}(h^{*})=0)_{1\leq i\leq n} \]

we can have the new pair of conditions for maximizing $L$

\begin{eqnarray*}
(\frac{\delta L}{\delta h_i}(h^{*})=0, \lambda^{*}=0)_{1\leq i\leq n} \land \lambda_k \geq 0 \  \text{or} \\
(\frac{\delta L}{\delta h_i}(h^{*})=0)_{1\leq i\leq n} \land \sum_i h_i f_{i,k}\geq F_k \land \lambda_k =0
\end{eqnarray*}

We first consider the derivative $\frac{\delta L}{\delta h_i}(h^{*})$ because this is  the only derivative that needs to be satisfied. This process is the same as equality constraints.

So, the maximum can be found by solving for all $i$ 
\[\frac{\delta L(h_i)}{h_i} =0\]
Recall our previous analysis of the Lagrange function:

\begin{eqnarray*}
L(h_i)&=& I(h;O)+d\sum_k \lambda_k (\sum_i h_i f_{i,k}-F_k)\\
&=&  d \sum_{o_s\in \hat{O_i} } h_i\phi_{s,i} \ln({\frac{\phi_{s,i}}{o_s}}) + d \sum_{o_s\in \hat{O_i},h_r \in P_i}  h_r\phi_{s,r}\ln(\frac{\phi_{s,r}}{o_s}) + d\sum_k \lambda_k (\sum_i h_i f_{i,k}-F_k)
\end{eqnarray*}

We solve the derivatives for each item in the Lagrange function. For the first item:
\begin{eqnarray*}\label{fder}
\frac{\delta (d\sum_{o_s \in \hat{O_i} } h_i\phi_{s,i} \ln({\frac{\phi_{s,i}}{o_s})}) }{\delta h_i}&=&d\sum_{o_s\in \hat{O_i}} \phi_{s,i} \ln( \frac{\phi_{s,i} } { o_s} )  - \frac{\phi_{s,i}^2 h_i}{o_s}
\end{eqnarray*}

For the second item:
\begin{eqnarray*}\label{sder}
\frac{\delta (d\sum_{o_s\in \hat{O_i},h_r \in P_i}  h_r\phi_{s,r}\ln(\frac{\phi_{s,r}}{o_s}) ) }{\delta h_i}&=&\\
\frac{\delta(d\sum_{o_s\in \hat{O_i},h_r \in P_i}  h_r\phi_{s,r}\ln \phi_{s,r}-\sum_{o_s\in \hat{O_i},h_r \in P_i}  h_r\phi_{s,r}\ln{o_s})  }{\delta h_i}&=&\\
0-d\sum_{o_s\in \hat{O_i},h_r \in P_i}  h_r\phi_{s,r}\frac{\phi_{s,i}}{o_s}&=&\\
-d\sum_{o_s\in \hat{O_i},h_r \in P_i}  h_r\phi_{s,r}\frac{\phi_{s,i}}{o_s}&&
\end{eqnarray*}

Because for $h_r\in P_i$,  $h_r\phi_{s,r}\ln \phi_{s,r}$ does not include any $h_i$, then the derivative by $h_i$ is 0. 

We combine the first two items and then simplify the expression as follows:
\begin{eqnarray*}\label{non-det-s1}
d\sum_{o_s\in \hat{O_i}} \phi_{s,i} \ln( \frac{\phi_{s,i} } { o_s} )  -d \frac{\phi_{s,i}^2 h_i}{o_s}  -d\sum_{o_s\in \hat{O_i},h_r \in P_i}  h_r\phi_{s,r}\frac{\phi_{s,i}}{o_s}&=&\\
d\sum_{o_s\in \hat{O_i}} \phi_{s,i} \ln( \frac{\phi_{s,i} } { o_s} )    -d\sum_{o_s\in \hat{O_i},h_r \in P_i}   \phi_{s,i}(\frac{\phi_{s,i} h_i}{o_s}+\frac{h_r\phi_{s,r}}{o_s})&=&\\
d\sum_{o_s\in \hat{O_i}} \phi_{s,i} \ln( \frac{\phi_{s,i} } { o_s} )    -d\sum_{o_s\in \hat{O_i}}   \phi_{s,i}(\frac{o_s}{o_s})&=&\\
d\sum_{o_s\in \hat{O_i}} (\phi_{s,i} \ln( \frac{\phi_{s,i} } { o_s} )    -  \phi_{s,i})&=&\\
d(\sum_{o_s\in \hat{O_i}} \phi_{s,i} \ln( \frac{\phi_{s,i} } { o_s} )    -  1)&&
\end{eqnarray*}

For the third item, the result is a linear function of $h_i$:
\begin{eqnarray*}\label{tder}
\frac{\delta (d\sum_k \lambda_k (\sum_i h_i f_{i,k}-F_k) ) }{\delta h_i}=d\sum_k \lambda_k f_{i,k}
\end{eqnarray*}

From these results we conclude that $max I(h; O)$ can be achieved by solving $h_i$ in the following equation system:
\begin{eqnarray*}\label{non-det}
\frac{\delta(L(h_i))}{h_i}=0 \Rightarrow  &&\\
d(\sum_{o_s\in \hat{O_i}} \phi_{s,i} \ln( \frac{\phi_{s,i} } { o_s} )    -  1)+d\sum_k \lambda_k f_{i,k}=0  \Rightarrow  &&\\
\sum_{o_s\in \hat{O_i}} \phi_{s,i} \ln( \frac{\phi_{s,i} } { o_s} )    -  1+\sum_k \lambda_k f_{i,k}=0
\end{eqnarray*}

As mentioned before, this equation needs to satisfy the following condition:

\begin{eqnarray*}
\lambda_k \geq 0 \ \text{or} \ \ \sum_i h_i f_{i,k} \geq F_k \land \lambda_k =0
\end{eqnarray*} 

when $\lambda_k=0$, the equation can be simplified to 

\[\sum_{o_s\in \hat{O_i}} \phi_{s,i} \ln( \frac{\phi_{s,i} } { o_s} )    -  1+\lambda_0=0 \]

where $\lambda_0$ is for the constraint $\sum_{i}h_i=1$.

So, we arrive at the conclusion that to maximize $L$, the following equations need to be solved with the constraints:
\begin{eqnarray*}
\sum_{o_s\in \hat{O_i}} \phi_{s,i} \ln( \frac{\phi_{s,i} } { o_s} )    -  1+\sum_k \lambda_k f_{i,k}=0 \land \lambda_k \geq 0 \ \text{or} \\
\sum_{o_s\in \hat{O_i}} \phi_{s,i} \ln( \frac{\phi_{s,i} } { o_s} )    -  1+\lambda_0=0 \land \sum_i h_i f_{i,k} \geq F_k
\end{eqnarray*} 

The proof completes.
}
\vspace{-1em}

If the system is completely deterministic, that is one input can only generate one ``observation", then the $o_j$'s are defined in terms of the high inputs $h_{j_i}$ that generate the ``observation", i.e.  
\[ o_j=h_{j_1}+\dots+h_{j_n} \]
Notice then that $(h_i|o_j)=\frac{(h_i,o_j)}{o_j}$  and that 
$(h_i,o_j) = h_i\ \ \text{if } h_i  \text{ generates the observation } o_j  \text{ otherwise is }   0$.

Because there is only one possible observation in the model associated with a high input $h=v_i$; denoted as $O(h_i)$ and
defined as $\hat{o_i} =\mu(O(h_i))$ 

Hence, we can simplify the Theorem \ref{non-det-ieq} to the following proposition by replacing $\phi_{i,s}$ with $1$:
\begin{proposition}\label{det-ieq}
In deterministic channels, the probabilities $h_i$ maximizing $I(h; O) $ subject to the family of constraint $(\mathcal{C}_k)_{k \in K} \equiv g_k(h_{i}) \geq F_k$ are given
 by solving in $h_i$ the following system of inequalities:
\[ -  \ln(  o_s )    -  1+ \sum_k \lambda_k f_{i,k}=0 \land  \lambda \geq 0 \land g_k(h_{i}) \ge F_k \]
or
 \begin{eqnarray*}
 - \ln( o_s )    -  1+\lambda_0&=&0 \land g_k(h_{i}) \ge F_k
\end{eqnarray*}
\end{proposition}

\begin{proposition}\label{Kmaxleakage:chanCap}
In both probabilistic and deterministic channels, the channel capacity without given knowledge is given by
\[ \sum_i h_i(1- \sum_k \lambda_kf_{i,k})d\]
In the case of $\lambda_k=0$, for all $k>0$  that simplifies to
\[  d(1-\lambda_0) \]
where $d=\frac{1}{\ln 2}$.
\end{proposition}

\myproof{
\begin{eqnarray*}
H(h)-H(h \vert O) &=&H(h) +  \sum_{j,i} \phi_{j,i} h_i\log(\frac{\phi_{j,i} h_i}{o_j})\\
&=&H(h)- \sum_{j,i} \phi_{j,i} h_i\log(\frac{o_j}{\phi_{j,i}}) +\sum_{j,i} \phi_{j,i} h_i\log(h_i)\\
&=&H(h)- \sum_{j,i} \phi_{j,i} h_i\log(\frac{o_j}{\phi_{j,i}}) -H(h)\\
&=&- \sum_{j,i} \phi_{j,i} h_i\log(\frac{o_j}{\phi_{j,i}}) \\
&=& \sum_{i} h_i \sum_{j} \phi_{j,i} \log(\frac{\phi_{j,i}}{o_j}) \\
&=& \sum_{i} h_i(   1- \sum_k \lambda_k f_{i,k})d
\end{eqnarray*}

where in deterministic channels $\phi_{j, i}=1$.
In the case $\lambda_k=0 (k \geq1)$ the expression becomes:
\[ d \sum_i h_i (1- \lambda_0)\]
which indicates one possible result.\\
The proof completes.
}
\vspace{-2em}
\subsection{Comparison with The Results Using Lagrange Multipliers}
From Theorem \ref{non-det-ieq} we notice that, in the solution of a constrained optimization problem, the inequality constraints either constrain the solution (i.e. $\lambda_i \neq 0 \land g_k(h_i)=F_k$), or they do not (i.e. $\lambda_i = 0 $). If they do, we can use Lagrange Multiplier to find the optimal solution by treating the inequality constraints as equality ones; otherwise, the constraints do not affect the solution. So, does it mean that the channel capacity theorem deduced by KKT has no improvement upon \cite{3han, 3pas}? The answer is no, because when there is a set of inequality constraints, it is difficult to determine which of them are constraining the problem. Then a method of classification is necessary to check whether the inequality constraints constrain or not. This is exactly what KKT conditions are doing: whether the constraints constrain the maxima or not, KKT deals with them elegantly.


\section{Applications of the Results}\label{app}
Theorem \ref{non-det-ieq} and Proposition \ref{Kmaxleakage:chanCap} can be applied in both programs and protocols to solve channel capacity with inequality constraints. 
In this section, two examples (a program and a protocol) will be studied to show how Theorem \ref{non-det-ieq} and Proposition \ref{Kmaxleakage:chanCap} are applied. 
The results are explained. 
Further, a short discussion is given on implementing this approach for automatic computation.

\subsection{Example: A Multi-threaded Program}
Let us start with a simple probabilistic nested multi-threaded program:
\begin{verbatim} 
l=h % 2 | (l=0 |  l=1) 
\end{verbatim}
Suppose that the outer thread has probability $p$ to run first ``l=h \% 2" and the inner thread has probability $q$ to run ``l=0" before  ``l=1" . From the program we know that there are two possible observations: $0(O_0)$ and $1(O_1)$.
We list all the possible values of $h$, observations and the conditional probabilities in Table \ref{pp1}.
 
\begin{table}
\small
\begin{center}
\begin{tabular}{r|r|r}
\hline
$h$ & $\phi_{( O_0, h)} $  &  $\phi_{( O_1,h)}$\\
\hline
$h_{odd}$ & $p(1-q)+(1-p)(1-q)p$  & $1-p(1-q)-(1-p)(1-q)p$\\
\hline
$h_{even} $ & $1-pq-(1-p)pq$  & $pq+(1-p)pq$\\
\hline
\end{tabular}
\end{center}
\caption{A multi-threaded program: observations and probabilities}
\label{pp1}
\end{table}
Assume $h$ is strictly less likely to be odd than even, i.e. the constraint on the input is:
\[h_{odd} <  h_{even} \]
Using Theorem \ref{non-det-ieq} we get equations:
\begin{eqnarray*}
-a\ln(\frac{ah_{odd}+bh_{even}}{a})-(1-a)\ln(\frac{(1-a)h_{odd}+(1-b)h_{even}}{1-a})  -  1+ \lambda_0+\lambda_1&=&0\\
-b\ln(\frac{ah_{odd}+bh_{even}}{b})-(1-b)\ln(\frac{(1-a)h_{odd}+(1-b)h_{even}}{1-b}) -  1+ \lambda_0-\lambda_1&=&0
\end{eqnarray*}
where $a=p(1-q)+(1-p)(1-q)p$ ; $\ b=1-pq-(1-p)pq$.\\
Firstly we consider the extreme case $p=1$  and we solve the equation system to get
\[\lambda_0=1, \ \lambda_1=0\]
Using Proposition \ref{Kmaxleakage:chanCap} we know that the channel capacity is $0$. This is because when $p=1$ which means ``l=h \% 2" running first then the program is secure because the result can not reveal any information of the secret.
Now we suppose $p=q=\frac{1}{3}$, and according to that we can solve $a=0.3704 \ \ b=0.8148$. 
Because the inequality is strict, we cannot have $h_{odd}=h_{even}$.
Thus we consider if the other possibility in Theorem \ref{non-det-ieq} $\lambda_1=0$ can be satisfied and we find:
\[ h_{odd}=0.4836 , h_{even}=0.5164, \lambda_0=0.8931, \lambda_1=0\]
This solution does satisfy $ h_{odd} < h_{even}$ and the distribution is the one we are after\footnote{Notice that values of $ h_{odd}>0.4836$ results in a lower leakage}.
Using Proposition \ref{Kmaxleakage:chanCap} we get the channel capacity:
\[  d(h_{odd}(1-\lambda_0-\lambda_1)+h_{even}(1-\lambda_0+\lambda_1)) =  0.1069 \ \textrm{bits} \]
The channel capacity is small, because among the three statements, only when ``l = h \% 2" is run in the end the program leaks, and the leakage is $1$ bit. The other two statements do not contribute to the leakage but further confuse the observation by producing same outputs $0$ and $1$, making the leakage even smaller.


\subsection{Example: Onion Routing}\label{ORSec}
 Onion Routing \cite{1ree} is designed to protect data and sender anonymity in 
communication over a public network such as the Internet.
The general idea is, when a client (sender) wants to send a message to a receiver $r$,
it will choose a path $p_1,\dots,p_n$ of routers and encrypt the message $m$ as $P_1(\dots(P_n(R(m),r))\dots, 2)$ where $P_i$ (resp $R$) is the public key of the router $i$ (resp receiver $r$). When the router $p_i$ receives $P_i(\dots(P_n(R(m),r))\dots, i+1)$ it will uses its private key to decrypt 
the message and will so get \\
$P_{i+1}(\dots(P_n(R(m),r))\dots, i+1)$, so it will send the message $P_{i+1}(\dots(P_n(R(m),r))\dots, i+2)$ to $p_{i+1}$.
Usual assumptions are:
\begin {enumerate}
\item A circuit can be of any number of nodes as long as no node appears twice. 
\item The client never sends the message to the server directly.
\item Observations of a node include the previous node and the next one.
\item All paths are equally likely.
\end {enumerate}

If the attacker can observe  one router $p_i$  then there may be a loss of anonymity: the attacker is able to observe which node delivered the packet to it and which node the packet is then be delivered to.

\begin{figure}[htp]
\centering
\includegraphics[width=0.4\textwidth]{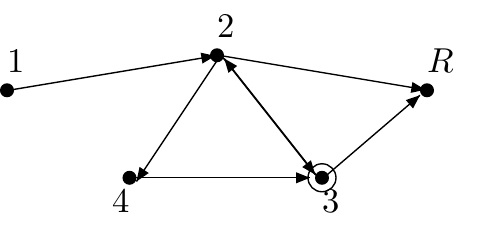}\\	
\caption{Example of An Onion Routing Network} 
\label{fig:on1} 
\end{figure}

Here we will show how the loss of sender anonymity can be quantitatively analyzed using the definition of channel 
capacity. We use the same simple Onion Routing network from \cite{3han} as shown in 
Figure \ref{fig:on1} but different and meaningful constraints will be demonstrated. The node ``R" is the receiver. There are 4 nodes 1,2,3,4 
in which either of them can initiate the communication; node 3 is an adversary
in the network. We list all the possible paths, observations on the adversary node 
and the conditional probabilities for the observations in the Table \ref{Top1}.

\begin{table}
\small
\begin{center}
\begin{tabular}{r|r|r|r}

\hline

$h$ & Path & O (in, out)  & $\phi_{O (in, out), h_i}$\\

\hline

1($h_1$)  &  $1\rightarrow 2 \rightarrow R$ & (N, N) & $\frac{1}{3}$\\
  &  $1\rightarrow 2 \rightarrow 3 \rightarrow R$ & (2, R) & $\frac{1}{3}$\\
    & $ 1\rightarrow 2 \rightarrow 4 \rightarrow 3 \rightarrow R$ & (4, R) & $\frac{1}{3}$\\
2($h_2$)  &  $2 \rightarrow 4 \rightarrow 3 \rightarrow R$ & (4, R) & $\frac{1}{2}$\\
  &  $2\rightarrow 3  \rightarrow R$ & (2, R) & $\frac{1}{2}$\\
3($h_3$)  &  $3 \rightarrow 2  \rightarrow R$ & (N, R) &$1$\\
4($h_4$) &  $4 \rightarrow 3 \rightarrow R$ & (4, R) & $\frac{1}{2}$\\
  &  $4\rightarrow 3  \rightarrow 2 \rightarrow R$ & (4, 2) & $\frac{1}{2}$\\
\hline     

\end{tabular}
\end{center}
\caption{Onion Routing: observations and probabilities}
\label{Top1}
\end{table}

From the Table \ref{Top1}, we get $o$ using  $o_j=\sum_{i}\phi_{j,i}$:
\[ o_{(N, N)}=\frac{1}{3}h_1,\  \ o_{(2, R)}=\frac{1}{3}h_1+\frac{1}{2}h_2, \] 
\[ o_{(4, R)}=\frac{1}{3}h_1+\frac{1}{2}h_2+\frac{1}{2}h_4, \ \ o_{(N,R)}=h_3, \  \ o_{(4,2)}=\frac{1}{2}h_4 \]

We now consider the case when an active user sends out messages more frequently than non-active users. Here we assume $h_1$ has greater probability than the node $h_2$. Then we have an additional constraint 
$h_1 \geq h_2$
with the constraint $C_0: h_1+h_2+h_3+h_4=1$.

We use Theorem \ref{non-det-ieq} to get the following equations:
\begin{align*}
-\frac{1}{3}( \ln(\frac{o_{(N, N)}}{\frac{1}{3}})+\ln(\frac{o_{(2, R)}}{\frac{1}{3}})+\ln(\frac{o_{(4, R)}}{\frac{1}{3}}))-1+\lambda_0+\lambda_1=0\\
-\frac{1}{2}\ln(\frac{o_{(2, R)}}{\frac{1}{2}})-\frac{1}{2}\ln(\frac{o_{(4, R)}}{\frac{1}{2}})-1+\lambda_0 - \lambda_1=0\\
-\ln o_{(N,R)} -1 +\lambda_0=0\\
-\frac{1}{2}\ln(\frac{o_{(4,2)}}{\frac{1}{2}})-\frac{1}{2}\ln(\frac{o_{(4, R)}}{\frac{1}{2}})-1+\lambda_0=0
\end{align*}

We firstly consider if the equality $h_1=h_2$ satisfies, then we solve the above equations and we find 
\[ h_1=0.1674 , h_2=0.1674, h_3=0.3903, h_4=0.2750, \]
\[\lambda_0=0.0591 , \lambda_1=-0.0072\]
But this solution does not satisfy $\lambda \geq 0$.

Then we only consider the solution for the other possibility $\lambda_1=0$, and we get the results:
\[ h_1=0.1735 , h_2=0.1603, h_3=0.3902, h_4=0.2760, \]
\[\lambda_0=0.0590 , \lambda_1=0\]

This solution does satisfy $h_1>h_2$. Using Proposition \ref{Kmaxleakage:chanCap} we get the channel capacity:
\[  d(h_1(1-\lambda_0-\lambda_1)+h_2(1-\lambda_0+\lambda_1)+(h_3+h_4)(1-\lambda_0)) = 1.3576 \ \textrm{bits} \]

When we have a strict inequality constraint, as we mentioned before, it may find an approximate solution in case if the accurate solution can not be achieved. The following example shows such a case. 
Here we use a similar constraint as above, assuming that the first node is 100 times likely to send the message compared to the second:
\[ h_1 >100 h_2 \]
Using Theorem \ref{non-det-ieq} we know that the second equation above becomes
\begin{align*}
-\frac{1}{2}\ln(\frac{o_{(2, R)}}{\frac{1}{2}})-\frac{1}{2}\ln(\frac{o_{(4, R)}}{\frac{1}{2}})-1+\lambda_0 - 100\lambda_1=0
\end{align*}
while the other three equations stay the same because the change of constraint does not affect them.
From the above result we can know that the solution for the case $\lambda_0=0$ does not satisfy $h_1 > 100h_2 $. We can use the equality constraint instead to find an approximate solution.
Assuming $h_1 = 100h_2$, we have
\[ h_1=0.2868 , h_2=0.0029, h_3=0.3979, h_4=0.3125, \]
\[\lambda_0=0.0783 , \lambda_1=0.0024\]
Using Proposition \ref{Kmaxleakage:chanCap} we get the channel capacity:
\[  d(h_1(1-\lambda_0-\lambda_1)+h_2(1-\lambda_0+100\lambda_1)+(h_3+h_4)(1-\lambda_0)) =1.3297 \ \textrm{bits} \]

Note that this is an approximate solution achieved when $100h_2+\xi = h_1$, $\xi \rightarrow 0 $.


In the first case when the constraint is $h_1\geq h_2$, the channel capacity is $1.3576$ bits. We compute the original secret of  $1.9042$ bits, which means the protocol leaks up to $72.2\%$ confidential information. In the second case, where the constraint is $h_1>100h_2$, the channel capacity is $1.3297$ bits.  Since the original confidential information is $1.5946$ bits, the rate is increased to $83.4\%$ which means the system is much more insecure. The reason is, $h_1$ and $h_2$ share the same observations as $(4, R)$ and $(2, R)$. Once the attacker observers these pairs, he/she has can more confidently guess the initial sender to be $h_1$ than $h_2$ with knowledge of the constraint $h_1>100h_2$. Thus, the constraint does affect the security of the protocol by reducing the confusion between $h_1$ and $h_2$.

In both cases, the channel capacity is around $1.3$ bits, which seems to imply that the protocol is insecure. 
Two observations are in order. First notice that by repeating observations on these networks the loss of anonymity is not increased.
Secondly in the real deployment of onion routing on the Internet (such as Tor), there are hundreds of nodes, with complex connectivity frequently updated; because of the number of possible connections  in such large scale networks the channel capacity is very low.
 
We have only one constraint in the above cases, but from the formula in Theorem \ref{non-det-ieq}
\[ \sum_{o_s\in \hat{O_i}} \phi_{s,i} \ln( \frac{\phi_{s,i} } { o_s} )   -  1+ \sum_k \lambda_k f_{i,k}=0 \]
\noindent multiple constraints will only affect the last item $ \sum_k \lambda_k f_{i,k}$ in the equation system. The complexity is increased linearly by increasing the number of factors $\lambda_k$.

\subsection{A Note on Automatic Computation}

Automatic analysis of programs and protocols can be achieved in two steps. The first step is to analyze the program or protocol to deduce the statistical relationship between $O$ and $h$. Recent works to automate this part include \cite{1jon,kopf2} which tracks the analyzed program iteratively to derive a precise answer. Alternatively, \cite{1cho,1dan} used simulations to derive an estimation. For the particular example of anonymity routing protocols, it is also possible to work out the statistical relationship based on the graph topology including vertexes, edges and adversaries. Based on the relationship, the equation system can be produced using Theorem \ref{non-det-ieq}. The second step is the automatic solution of the equation system. Automated solution of such an equation system has been implemented in standard mathematical packages, e.g. MATLAB.

\section{Conclusion and Future Work}\label{clf}
We apply Karush-Kuhn-Tucker conditions to solve the channel capacity of probabilistic information leakage channels with inequality constraints. 
We derived a series of theorems and propositions and we show how these results can be applied to programs and protocols. 
Our calculations provide general and accurate solutions to measure the maximum information leakage in a system.

Our future work will investigate other continuous definitions of information leakage 
using Karush-Kuhn-Tucker conditions. Notably, we propose to solve the maximum ratio between the channel capacity of a leakage channel and that of the original secret,
which in some cases could present a better definition of {\it the worst case}. Additionally, 
a comparison of the information theoretical and probabilistic analysis of probabilistic channels \cite{FeJoSy,DiSeClPr} would also yield interesting results.

\end{document}